\documentclass[12pt]{article}
\usepackage{graphicx}
\begin{document}

\catcode`@=11
\long\def\@caption#1[#2]#3{\par\addcontentsline{\csname
  ext@#1\endcsname}{#1}{\protect\numberline{\csname
  the#1\endcsname}{\ignorespaces #2}}\begingroup
    \small
    \@parboxrestore
    \@makecaption{\csname fnum@#1\endcsname}{\ignorespaces #3}\par
  \endgroup}
\catcode`@=12
\newcommand{\newc}{\newcommand}
\newc{\lat}{{\ell at}}
\newc{\one}{{\bf 1}}
\newc{\mgut}{M_{\rm GUT}}
\newc{\mzero}{m_0}
\newc{\mhalf}{M_{1/2}}
\newc{\five}{{\bf 5}}
\newc{\fivebar}{{\bf\bar 5}}
\newc{\ten}{{\bf 10}}
\newc{\tenbar}{{\bf\bar{10}}}
\newc{\sixteen}{{\bf 16}}
\newc{\sixteenbar}{{\bf\bar{16}}}
\newc{\gsim}{\lower.7ex\hbox{$\;\stackrel{\textstyle>}{\sim}\;$}}
\newc{\lsim}{\lower.7ex\hbox{$\;\stackrel{\textstyle<}{\sim}\;$}}
\newc{\gev}{\,{\rm GeV}}
\newc{\mev}{\,{\rm MeV}}
\newc{\ev}{\,{\rm eV}}
\newc{\kev}{\,{\rm keV}}
\newc{\tev}{\,{\rm TeV}}
\newc{\mz}{m_Z}
\newc{\mw}{m_W}
\newc{\mpl}{M_{Pl}}
\newc{\mh}{m_h}
\newc{\mA}{m_A}
\newc{\tr}{\mbox{Tr}}
\def\sfrac#1#2{{\textstyle\frac{#1}{#2}}}
\newc{\chifc}{\chi_{{}_{\!F\!C}}}
\newc\order{{\cal O}}
\newc\CO{\order}
\newc\CL{{\cal L}}
\newc\CY{{\cal Y}}
\newc\CH{{\cal H}}
\newc\CM{{\cal M}}
\newc\CF{{\cal F}}
\newc\CD{{\cal D}}
\newc\CN{{\cal N}}
\newc{\eps}{\epsilon}
\newc{\re}{\mbox{Re}\,}
\newc{\im}{\mbox{Im}\,}
\newc{\invpb}{\,\mbox{pb}^{-1}}
\newc{\invfb}{\,\mbox{fb}^{-1}}
\newc{\yddiag}{{\bf D}}
\newc{\yddiagd}{{\bf D^\dagger}}
\newc{\yudiag}{{\bf U}}
\newc{\yudiagd}{{\bf U^\dagger}}
\newc{\yd}{{\bf Y_D}}
\newc{\ydd}{{\bf Y_D^\dagger}}
\newc{\yu}{{\bf Y_U}}
\newc{\yud}{{\bf Y_U^\dagger}}
\newc{\ckm}{{\bf V}}
\newc{\ckmd}{{\bf V^\dagger}}
\newc{\ckmz}{{\bf V^0}}
\newc{\ckmzd}{{\bf V^{0\dagger}}}
\newc{\X}{{\bf X}}
\newc{\bbbar}{B^0-\bar B^0}
\def\bra#1{\left\langle #1 \right|}
\def\ket#1{\left| #1 \right\rangle}
\newc{\sgn}{\mbox{sgn}\,}
\newc{\m}{{\bf m}}
\newc{\msusy}{M_{\rm SUSY}}
\newc{\munif}{M_{\rm unif}}
\newc{\slepton}{{\tilde\ell}}
\newc{\Slepton}{{\tilde L}}
\newc{\sneutrino}{{\tilde\nu}}
\newc{\selectron}{{\tilde e}}
\newc{\stau}{{\tilde\tau}}
\newc{\vbb}{\beta\beta 0\nu}
\newc{\vckm}{V_{\!\mbox{\tiny CKM}}}
\newc{\mbbeff}{m_{\beta\beta}^{\mbox{\tiny eff}}}
%
%
\def\NPB#1#2#3{Nucl. Phys. {\bf B#1} (19#2) #3}
\def\PLB#1#2#3{Phys. Lett. {\bf B#1} (19#2) #3}
\def\PLBold#1#2#3{Phys. Lett. {\bf#1B} (19#2) #3}
\def\PRD#1#2#3{Phys. Rev. {\bf D#1} (19#2) #3}
\def\PRL#1#2#3{Phys. Rev. Lett. {\bf#1} (19#2) #3}
\def\PRT#1#2#3{Phys. Rep. {\bf#1} (19#2) #3}
\def\ARAA#1#2#3{Ann. Rev. Astron. Astrophys. {\bf#1} (19#2) #3}
\def\ARNP#1#2#3{Ann. Rev. Nucl. Part. Sci. {\bf#1} (19#2) #3}
\def\MPL#1#2#3{Mod. Phys. Lett. {\bf #1} (19#2) #3}
\def\ZPC#1#2#3{Zeit. f\"ur Physik {\bf C#1} (19#2) #3}
\def\APJ#1#2#3{Ap. J. {\bf #1} (19#2) #3}
\def\AP#1#2#3{{Ann. Phys. } {\bf #1} (19#2) #3}
\def\RMP#1#2#3{{Rev. Mod. Phys. } {\bf #1} (19#2) #3}
\def\CMP#1#2#3{{Comm. Math. Phys. } {\bf #1} (19#2) #3}
\relax
%
%
%
\def\beq{\begin{equation}}
\def\eeq{\end{equation}}
\def\bea{\begin{eqnarray}}
\def\eea{\end{eqnarray}}
%
%
%
\newc{\ie}{{\it i.e.}}          \newc{\etal}{{\it et al.}}
\newc{\eg}{{\it e.g.}}          \newc{\etc}{{\it etc.}}
\newc{\cf}{{\it c.f.}}
\def\smuon{{\tilde\mu}}
\def\neut{{\tilde N}}
\def\char{{\tilde C}}
\def\bino{{\tilde B}}
\def\wino{{\tilde W}}
\def\higgsino{{\tilde H}}
\def\sneut{{\tilde\nu}}
%
%
%
%
\def\slash#1{\rlap{$#1$}/} 
\def\Dsl{\,\raise.15ex\hbox{/}\mkern-13.5mu D} 
\def\delsl{\raise.15ex\hbox{/}\kern-.57em\partial}
\def\Ksl{\hbox{/\kern-.6000em\rm K}}
\def\Asl{\hbox{/\kern-.6500em \rm A}}
\def\Qsl{\hbox{/\kern-.6000em\rm Q}}
\def\gradsl{\hbox{/\kern-.6500em$\nabla$}}
%
%
%
\def\bar#1{\overline{#1}}
\def\vev#1{\left\langle #1 \right\rangle}
%

\begin{titlepage}
\begin{flushright}
October 2008\\
\end{flushright}
\vskip 2cm
\begin{center}
{\large\bf
The Effect of Quark Sector Minimal Flavor Violation on \\ Neutrinoless Double Beta Decay}
\vskip 1cm
{\normalsize\bf
Brian Dudley and Christopher Kolda\\
\vskip 0.5cm
{\it Department of Physics, University of Notre Dame\\
Notre Dame, IN~~46556, USA}\\[0.1truecm]
}

\end{center}
\vskip .5cm

\begin{abstract}
The question of whether neutrino masses are Dirac or Majorana is one of the most important,
and most difficult, questions remaining in the neutrino sector. Searches for neutrinoless double $\beta$-decay may help to resolve this question, but are also sensitive to new, higher dimension $\Delta L=2$ operators. In this paper we place two phenomenological constraints on these operators at dimension $d\leq 11$. First, we require that the operators
obey the quark flavor symmetries of the Standard Model, with any violation of the symmetries being due to Yukawa interactions, a scheme known as Minimal Flavor Violation (MFV). Second, we require that the operators which generate neutrinoless double $\beta$-decay, and any operators related by the flavor symmetries, do not induce neutrino masses above $0.05\ev$, the limit implied by the atmospheric neutrino data.
We find that these requirements severely constrain the operators which can violate lepton number, such that most can no longer contribute to neutrinoless double $\beta$-decay at observable rates.  It is noteworthy that quark flavor symmetries can play such a strong role in constraining new leptonic physics, even when that physics is not quark flavor changing.
Those few operators that can mimic a Majorana neutrino mass then appear with cutoffs below a TeV, and represent new physics which could be directly probed at the LHC or a future linear collider.
\end{abstract}

\end{titlepage}

\setcounter{footnote}{0}
\setcounter{page}{1}
\setcounter{section}{0}
\setcounter{subsection}{0}
\setcounter{subsubsection}{0}

\section{Philosophy}

While the Standard Model (SM) of particle physics may have many unanswered questions, one of the few for which solid experimental data exists is the issue of neutrino masses. Given the multiple observations of neutrino oscillations, in
solar~\cite{Fukuda:2001nk}, reactor~\cite{Apollonio:1999ae,Ahn:2002up}
and atmospheric~\cite{Fukuda:1998mi,Ahmad:2002jz}
neutrino experiments, it appears that at least two of the three neutrino
species have non-zero masses, though they are still quite small.

Within the confines of the SM there is only one way to give a neutrino a mass,
and that is through a Majorana mass term, of the form $m_{\nu}\bar\nu^c_L\nu_L$. Such
terms are allowed after electroweak symmetry breaking because the neutrino carries no gauge
charges under the resulting SU(3)$\times$U(1)$_{EM}$. Further, such a term is naturally
generated at dimension five (and therefore is naturally small) once we allow non-renormalizable/irrelevant
operators into our Lagrangian. The presence of such a mass term is a signal of the non-conservation
of lepton number ($L$), and we would expect to find a host of other non-renormalizable operators which
likewise violate $L$.

On the other hand, lepton number can be preserved if the mass is obtained by minimally
extending the SM to include a right-handed neutrino state. Such a state could participate
with the left-handed neutrino in a Dirac mass term of the form $m_{\nu}\bar{\nu}_R\nu_L$. In this
case there is no fundamental reason for the lightness of neutrinos apart from the smallness of the
underlying Yukawa coupling. While ``technically natural" (they are preserved by chiral symmetries), such
small Yukawa couplings strike many physicists as deeply unnatural in a more generic sense.
But because the $\nu_R$ can be assigned a lepton number, $L$ is conserved.

So the question of whether neutrino masses are Majorana or Dirac is ultimately a
question of whether $L$ is a true symmetry of the SM Lagrangian or not, at least at the
perturbative level. A Majorana mass is particularly interesting not only because it
violates $L$, but because it is intrinsically non-perturbative, thereby implying a cut-off
for the SM above which some new physics must be present. Any measurement of a Majorana neutrino
mass automatically leads to a range of possible cutoffs at which we might expect new physics to appear.

\begin{figure}[ht]
\centering
\label{vbbdiag}
\includegraphics[scale = 1.00]{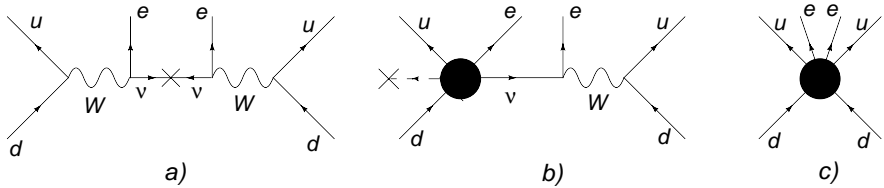}
\caption{Example Feynmann diagrams of $\vbb$ for a) dimension-$5$, b) dimension-$7$ and
c) dimension-$9$.}
\end{figure}

Unfortunately the question of whether the neutrino masses so far observed are Dirac
or Majorana is not a simple one. The only process by which we can reasonably expect to observe
the effects of a Majorana mass directly are the neutrinoless double $\beta$-decays ($\vbb$).
Normal double-$\beta$ decay occurs
in only a handful of nuclei, and represents the simultaneous $\beta$-decay of two neutrons, emitting in
the process two electrons and two anti-neutrinos. While rare, such a decay is allowed by the SM.
However if $L$ is violated and neutrinos are Majorana in character, the out-going anti-neutrino
from one $\beta$-decay can become an incoming neutrino for the second decay, thus eliminating both
final state neutrinos entirely; see Fig.~\ref{vbbdiag}(a).

The detection of this {\it neutrinoless}\/ process has driven a number of experiments in a variety of nuclei.
The most frequently studied nucleus is germanium-76, which has been used recently by the
Heidelberg-Moscow group~\cite{KlapdorKleingrothaus:2000sn,Baudis:1999xd}
and by IGEX~\cite{Aalseth:2002rf}. There are many excellent reviews discussing these and
other $\vbb$ experiments; see, for example, Ref.~\cite{Elliott:2002xe}. While there has been
one claim for a signal with a lifetime of approximately $10^{25}$ years~\cite{KlapdorKleingrothaus:2001ke}, it
has not been verified and is under contention~\cite{Aalseth:2002dt}.  A wide range of future experiments
will continue to search for this class of decays, pushing their sensitivity to smaller and smaller
neutrinos masses, closer to the values consistent with the oscillation experiments.

However there are several difficulties in interpreting any positive signal in $\vbb$ experiments,
above and beyond those of extracting and verifying the signal. First, a neutrino may have both Dirac
and Majorana masses and both may play important roles in the complete neutrino mass matrix. Second,
because neutrino weak eigenstates are not identical to their mass eigenstates, the mass
detected in $\vbb$ experiments
is not a physical mass; this is very well known and we will return to it briefly in the next
section. Finally, there may be other sources of $L$-violation besides Majorana neutrino masses, and
these new sources may be contributing to the observation of $\vbb$ with a greater strength than
that of the diagram in Fig.~\ref{vbbdiag}(a).

If other sources of $L$-violation are dominant in $\vbb$, then any interpretation of a signal
in terms of a Majorana neutrino mass will be flawed. This does not mean that observation of $\vbb$
cannot be interpreted as evidence for Majorana masses. In fact, Majorana masses
{\it are}\/ implied by observation of $\vbb$. Neutrinoless double-$\beta$ decay
is a $\Delta L=2$ process whose presence will lead to the generation of all other $\Delta L=2$ operators
in the SM Lagrangian, including the ($\Delta L=2$) neutrino Majorana mass term. But
one cannot simply assume that the Majorana mass can be cleanly extracted from a measurement
of the $\vbb$ lifetime because of the possibility of other $L$-violating operators.

For the purpose of this paper, we will refer to these other $L$-violating operators as ``new physics"
because they are all non-renormalizable and therefore provide a cut-off, with a requisite
scale $\Lambda$, for the SM. Above the scale $\Lambda$ some new physics must come in, new
physics that will violate $L$ at the renormalizable level.

There are in principle two ways to determine whether the dominant contribution to a $\vbb$ signal
is a neutrino mass or a new physics operator. First, one would attempt to
observe $\vbb$ in several different nuclei, using differing nuclear matrix elements
to tease out information on the underlying hard amplitudes. This is a difficult process,
both experimentally (requiring observation of $\vbb$ in multiple nuclei) and
theoretically (understanding all the relevant nuclear matrix elements at the required
precision, and accounting for all neutrino flavor mixing effects).

The second route is more straightforward and certainly more relevant as we enter the era of
the Large Hadron Collider (LHC). The LHC, and any future linear electron-positron collider,  will
be probing physics at roughly the 1~TeV scale. Were there new sources of $L$-violation at or
below the TeV scale, one would expect to see them~\footnote{There are of course some questions
of how to study leptonic physics at a hadron machine. Typically one does so by observing new
leptonic states or interactions in the cascade decays of any new strongly-interacting states. The
planned searches for selectrons from squark decays at the LHC is a good example. However it is possible
that new physics in the lepton sector could hide from a hadron collider, necessitating
studies using electrons.}. The relevant question, then, is: at what scale must new physics
lie if a signal is observed in $\vbb$?

In the next section, we will see that a naive analysis yields a
scale that is surprisingly high. For example, a signal in the present generation of $\vbb$ experiments
would indicate new physics at scales as high as several hundred TeV! While these
numbers mean that $\vbb$ experiments are probing scales far above those the LHC is capable
of studying (as has been advertised for this process), they also mean that no
planned collider will be capable of ruling out a new physics explanation for a $\vbb$ signal.

In this paper, we will argue that $\vbb$ is unlikely to probe such high scales, and that,
in fact, the current and planned generations of $\vbb$ experiments are probing scales much closer to
those being studied by the LHC. By ``unlikely" we mean that in well-motivated extensions
of the SM, there are large suppressions on the new physics operators which generate $\vbb$,
causing their effective cut-off scale to drop below a TeV. These suppressions have nothing at
all to do with the lepton sector of the theory, but are a direct consequence of the constraints
and symmetries in the quark sector!

In particular we will show that within the broad class of new physics models which exhibit
the property of ``minimal flavor violation"
(MFV)\cite{D'Ambrosio:2002ex}, new physics contributions to $\vbb$ are highly
suppressed and will only be visible in next-generation $\vbb$ experiments if the new
physics sits below the TeV scale. As we will explain in Section 3, MFV is a property of
the quarks alone, but is sufficient to suppress the leading $L$-violating operators which
contribute to $\vbb$. This is both a surprise and a general statement with regard to
new physics: if new operators violate the approximate global symmetries of the quark flavor
sector, they may need to be highly suppressed in order to avoid generating large
flavor-changing neutral currents. This result even impacts physics in sectors that are seemingly
far-removed from quark flavor constraints, including operators which are quark flavor conserving,
such as $\vbb$.

In section~2 of this paper we will review the canonical discussion of $\vbb$, including
both the effects of Majorana neutrino masses, and of higher-dimension, new physics
operators. In section~3 we will introduce minimal flavor violation and examine its
effects on new physics contributions to $\vbb$. Finally, in section~4, we will
quickly discuss the effect of neutrino masses induced
by $\vbb$ operators, before concluding it in section~5

\section{Canonical Approach to Neutrinoless Double $\beta$-Decay}

At the renormalizable level, the SM allows no violation of total
lepton number, $L$. But because $L$ is anomalous, and because its violation
is in several ways indirectly tied to baryon number violation and the generation
of net baryon number in the universe, it is highly doubtful that this apparent $L$-conservation
is anything more than an accidental symmetry of the $d\leq4$ terms in the Lagrangian.
Thus $L$-violation should be considered one of the first ``expected surprises" to be discovered
once one probes the structure of the SM at higher and higher precision.

Consistent with the gauge symmetries of the SM, the first term which violates $L$ is the
dimension-5, $\Delta L=2$ operator:
\bea
{\cal L}_{d=5}& =&-\frac{\lambda_{5}}{\Lambda}(LH)(LH) + h.c. \nonumber\\
&=& \frac{-\lambda_{5}v^2}{\Lambda} \nu \nu + h.c.
\label{eq:dim5}
\eea
(Note that throughout this paper we will be writing our fermionic fields as
two-component, Weyl spinors, as is common in discussions of new physics. Thus $L$ above
is the left-handed lepton doublet and $H$ is the Higgs doublet. SU(3) contractions,
wherever applicable, are left implicit.)

Once the electroweak symmetry is broken, the operator in the first line of Eq.~(\ref{eq:dim5}) breaks
to the Majorana mass term in the second line. The implied neutrino mass is
then $m_\nu=\lambda_5 v^2/\Lambda$, with $v=246\gev$.  Assuming $\Lambda\gg v$, the
neutrino masses are naturally small, explaining the observed hierarchy between the charged lepton and
neutrino masses. Thus the lowest order effect of $L$-violation is in fact Majorana neutrino masses.

But observation of neutrino masses alone is not sufficient for proving the existence
of $L$-violation; the previously discussed Dirac masses could also be responsible for any observed
mass. In order to test for Majorana masses one needs an experiment capable of observing $L$-violation
directly. Within the framework of the SM, that experiment is neutrinoless double-$\beta$ decay.
Majorana neutrino masses contribute to $\vbb$ through the diagram in Fig.~\ref{vbbdiag}(a).
In order to complete the
internal neutrino line, a helicity flip is provided by the Majorana mass. Thus the rate for $\vbb$ in $^{76}$Ge is
proportional to the square of the Majorana mass:
\begin{equation}
\tau^{-1} = 4.6\times10^{-31}\left[\frac{m_{\beta\beta}}{1\times 10^{-3}\mbox{eV}}\right]^2 \mbox{ yr}^{-1}.
\label{eq:dim5rate}
\end{equation}
Because the weak currents couple only to the $\nu_e$ component of the physical neutrinos,
the neutrino mass appearing in the rate, $m_{\beta\beta}$ is a function of the physical masses and mixings:
\beq
m_{\beta\beta}=\sum_{i=1}^{3}\left(U_{ei}\right)^2 m_{\nu_i},
\label{eq:mnunu}
\eeq
where $U$ is the neutrino mixing matrix and $m_{\nu_i}$ are the physical neutrino masses
(assuming they are pure Majorana).

In Eq.~(\ref{eq:dim5rate}) we have normalized the rate assuming a neutrino mass of $1$ meV. So the controversial
signal of $\vbb$ of Ref.~\cite{KlapdorKleingrothaus:2001ke}
with a lifetime of $\tau\approx10^{25}\,$years corresponds
to a neutrino mass of about $0.5\ev$.  A neutrino mass of about $0.1\ev$ would
correspond to a lifetime of about $10^{26}\,$years, and so on. Note that such a small
neutrino mass requires either $\Lambda$ be very large, about
$10^{11}$ TeV, or alternatively, the coupling constant $\lambda_5$ to be very tiny.

If the scale $\Lambda$ of $L$-violation is in fact so much higher than the weak scale,
then the leading effect of $L$-violation will be Majorana neutrino masses, and the only
experiment for uncovering their presence will be $\vbb$.
But if the scale $\Lambda$ is relatively close to the weak scale ($\Lambda\lsim 10\tev$),
then two new possibilities open up.

First, high energy colliders could probe the physics of the scale $\Lambda$ directly,
finding and studying a new particle at that scale whose interactions do not conserve $L$. A good example
would be observation of supersymmetric partners along with $R$-parity violating interactions. Such findings
would be a valuable complement to the $\vbb$ experiments and would allow physicists to piece together a
much clearer picture of lepton number.

The second possibility is that new operators would contribute to $\vbb$, apart from the Majorana
mass term. Neutrinoless double-$\beta$ decay is itself a dimension-9 process represented by a set of
operators with the general form $(1/\Lambda^5)\bar{d}u\bar{d}uee$. If $\lambda_5$ is very small
then the direct dimension-9 operator for $\vbb$ may dominate
the experimental lifetime. In such a case, attempts to interpret a lifetime measurement in terms of a
neutrino mass would be fruitless.

There are also dimension-$7$ operators which, when dressed with weak interactions,
contribute to $\vbb$ through diagrams like Fig.\ref{vbbdiag}(b)~\cite{Babu:2001ex}. Compared to
operators which are intrinsically dimension-$9$, the
dimension-$7$ operators lead to rates for $\vbb$ enhanced by $(\Lambda^4/m_p^2v^2)$. At the opposite end, there are
dimension-$11$ operators which contain 6 fermion fields and have two Higgs
doublets, which allows for more complicated SU(2) contractions which than
can be obtained from dimension-$9$ operators alone.  Dimension-$11$ operators, like dimension-$9$, contribute
through the point-like diagrams seen in Fig.~\ref{vbbdiag}(c),
but the rates are further suppressed by $(v/\Lambda)^4$.  There are even higher dimension operators
suppressed by even more powers of $(v/\Lambda)^4$, but, we will ignore them in this paper.

\subsection{New Physics Contributions}

The operators which generate $\vbb$ at $d=7\mbox{, }9\mbox{ and }11$ have been studied previously
and we will only review them here. Throughout we will use Weyl notation, with $\sigma^{i}$ being
the Pauli matrices, $\sigma^{\mu} = (1,\sigma^i)$, $\bar{\sigma}^{\mu} = (1,-\sigma^i)$, and
$\sigma^{\mu\nu} = \frac{1}{4}(\sigma^{\mu}\bar{\sigma}^{\nu}-\sigma^{\nu}\bar{\sigma}^{\mu})$.
We will write the $\Delta L = 2$ operators in the form
$(\lambda_{i,j}/\Lambda^{i-4}){\cal O}_{i,j}$, where $\lambda_{i,j}$
represents the coupling constant for the $j$th operator of dimension $i$. We will explicitly
preserve the Lorentz structure in the hopes of better quantifying the
effects of the nuclear matrix elements.  We will return to this topic further in.

The dimension-$7$ operators that contribute to $\vbb$ are all of the
form $\bar{q}\bar{q}\ell\nu H$ where $\nu$ and sometimes to other fields
are members of LH weak doublets.  After electroweak symmetry breaking, the Higgs is replaced
by its vev, yielding the following operators:
\bea
{\cal L}_{d=7} & = & \frac{v}{\sqrt{2}\Lambda^3}\left[\lambda_{7,1} (\nu_e e)(u d^c) \right. \nonumber\\
&& \left. +\lambda_{7,2}(\nu_e \sigma^{\mu\nu}e)(u \sigma_{\mu\nu} d^c)+
\lambda_{7,3} (\nu_e e)(\bar{d} \bar{u}^c)\right.\nonumber\\
&&\left.+\lambda_{7,4}(\nu_e \sigma^{\mu\nu} e)(\bar{d} \sigma_{\mu\nu} \bar{u}^c)
+\lambda_{7,5}(\nu_e \sigma^{\mu} \bar{e}^c)(d^c \sigma_{\mu}\bar{u}^c)\right]+h.c.,
\eea
where charge conjugated fields, $u^c, d^c\mbox{ and }e^c$ are singlets under $\mbox{SU}(2)$. Fields
without the charge conjugation transform as doublets under $\mbox{SU}(2)$.  These operators
are all dressed by the weak interactopm; the neutrino becomes part of a weak current,
producing the dimension-$9$ operator of $\vbb$ as in Fig.~\ref{vbbdiag}(b).

The dimension-$9$ operators, which consist purely of short range physics, are given as follows:
\bea
{\cal L}_{d=9} & =& \frac{1}{\Lambda^5}\left[\lambda_{9,1}(e e)(u d^c)(u d^c)+
\lambda_{9,2}(e e)(u\sigma^{\mu\nu} d^c)(u\sigma_{\mu\nu} d^c)\right.\nonumber\\
&&\left.+\lambda_{9,3}(e e)(\bar{d} \bar{u}^c)(\bar{d} \bar{u}^c)+
\lambda_{9,4}(e e)(\bar{d}\sigma^{\mu\nu}\bar{u}^c)(\bar{d}\sigma_{\mu\nu} \bar{u}^c)\right.\nonumber\\
&&\left.+\lambda_{9,5}(e e)(u d^c)(\bar{d} \bar{u}^c)+
\lambda_{9,6}(e e)(u\sigma^{\mu\nu}d^c)(\bar{d}\sigma_{\mu\nu} \bar{u}^c)\right.\nonumber\\
&&\left.+\lambda_{9,7}(e e)(u\sigma^{\mu}\bar{d})(\bar{u}^c\sigma_{\mu}d^c)\right.\nonumber\\
&&\left.+\lambda_{9,8}(e\sigma^{\mu}\bar{e}^c)(d^c\sigma_{\mu}\bar{u}^c)(u d^c)+
\lambda_{9,9}(e\sigma^{\mu} e)(d^c\sigma^{\nu}\bar{u}^c)(u\sigma_{\mu\nu}d^c)\right.\nonumber\\
&&\left.+\lambda_{9,10}(e\sigma^{\mu}\bar{e}^c)(d^c\sigma_{\mu}\bar{u}^c)(\bar{d} \bar{u}^c)+
\lambda_{9,11}(e\sigma^{\mu} e)(d^c\sigma^{\nu}\bar{u}^c)(\bar{d}\sigma_{\mu\nu}\bar{u}^c)\right.\nonumber\\
&&\left.+\lambda_{9,12}(\bar{e}^c \bar{e}^c)(d^c\sigma_{\mu}\bar{u}^c)(d^c\sigma^{\mu}\bar{u}^c)\right]+h.c.
\eea

Finally, the dimension-$11$ operators are:
\bea
{\cal L}_{d=11}& =&\frac{v^2}{2\Lambda^7}\left[\lambda_{11,1}(e e)(u\sigma^{\mu}\bar{d})(u\sigma_{\mu}\bar{d})
+\lambda_{11,2}(e e)(d^c\sigma^{\mu}\bar{u}^c)(d^c\sigma_{\mu}\bar{u}^c)\right.\nonumber\\
&&\left.+\lambda_{11,3}(e\sigma^{\mu}\bar{e}^c)(u\sigma_{\mu}\bar{d})(u d^c)+
\lambda_{11,4}(e\sigma^{\mu}\bar{e}^c)(u\sigma^{\nu}\bar{d})(u\sigma_{\mu\nu}d^c)\right.\nonumber\\
&&\left.+\lambda_{11,5}(e\sigma^{\mu}\bar{e}^c)(u\sigma_{\mu}\bar{d})(\bar{d} \bar{u}^c)+
\lambda_{11,6}(e\sigma^{\mu}\bar{e}^c)(u\sigma^{\nu}\bar{d})(\bar{d}\sigma_{\mu\nu}\bar{u}^c)\right.\nonumber\\
&&\left.+\lambda_{11,7}(\bar{e}^c \bar{e}^c)(u d^c)(u d^c)+
\lambda_{11,8}(\bar{e}^c \bar{e}^c)(u\sigma^{\mu\nu}d^c)(u\sigma_{\mu\nu}d^c)\right.\nonumber\\
&&\left.+\lambda_{11,9}(\bar{e}^c \bar{e}^c)(\bar{d} \bar{u}^c)(\bar{d} \bar{u}^c)+
\lambda_{11,10}(\bar{e}^c \bar{e}^c)
(\bar{d}\sigma^{\mu\nu}\bar{u}^c)(\bar{d}\sigma_{\mu\nu}\bar{u}^c)\right.\nonumber\\
&&\left.+\lambda_{11,11}(\bar{e}^c \bar{e}^c)(u d^c)(\bar{d} \bar{u}^c)+
\lambda_{11,12}(\bar{e}^c \bar{e}^c)(u\sigma^{\mu\nu}d^c)(\bar{d}\sigma_{\mu\nu}\bar{u}^c)\right.\nonumber\\
&&\left.+\lambda_{11,13}(\bar{e}^c \bar{e}^c)(u\sigma^{\mu}\bar{d})(u\sigma_{\mu}\bar{d})\right]+h.c.
\eea
The original dimension-$11$ operators contain six fermion fields and two Higgs fields, where
$SU(2)$ indices of the Higgs Fields are contracted with those of the fermions. Dimension-$11$
operators in which the Higgas fields contract amongst themselves(\ie $H^i\bar{H}_i$) are not
included, since these are simply higher order corrections to existing dimension-$9$ operators.

There are other more extensive lists of lepton number violating (LNV) operators in~\cite{Babu:2001ex,de Gouvea:2007xp}.
The list we examine above differs from these in that we are interested
in operators capable of mediating $\vbb$ directly,
rather than generating $\vbb$ through an induced Majorana neutrino mass.  For this reason
our operator list is smaller than those found in the papers above. We also keep the explicit Lorentz structure of
each operator. However, the operators we use in this paper can be compared to the operators from the referenced papers.
For example, our dimension-$9$ operators correspond to the operators $11b, 12a, 14b, 19 \mbox{ and } 20$ in
the referenced papers.
\begin{table}[t]
\centering
\begin{tabular}{c|c|c|c}
$\lambda_{7,1},\lambda_{7,3}$&$\lambda_{7,2}$&$\lambda_{7,4}$&$\lambda_{7,5}$\\
\hline
$6.9\times10^{-10}$&$2.9\times10^{-8}$&$2.0\times10^{-7}$&$1.7\times10^{-13}$\\
\end{tabular}

\vspace{0.5cm}

\begin{tabular}{c|c|c|c|c|c}
$\lambda_{9,1},\lambda_{9,3},\lambda_{9,5}$&$\lambda_{9,2},\lambda_{9,4},\lambda_{9,6}$&
$\lambda_{9,7}$&$\lambda_{9,8},\lambda_{9,10}$&$\lambda_{9,9},\lambda_{9,11}$&$\lambda_{9,12}$\\
\hline
$6.2\times10^{-13}$&$1.4\times10^{-8}$&$5.6\times10^{-10}$&$1.4\times10^{-12}$&
$1.4\times10^{-10}$&$3.5\times10^{-11}$\\
\end{tabular}

\vspace{0.5cm}

\begin{tabular}{c|c|c|c|c|c}
$\lambda_{11,1},\lambda_{11,2}$&$\lambda_{11,3},\lambda_{11,5}$&$\lambda_{11,4},\lambda_{11,6}$&
$\lambda_{11,7},\lambda_{11,9},\lambda_{11,11}$&$\lambda_{11,8},\lambda_{11,10},\lambda_{11,12}$&$\lambda_{11,13}$\\
\hline
$3.5\times10^{-11}$&$1.4\times10^{-12}$&$1.4\times10^{-10}$&$6.2\times10^{-13}$&$1.4\times10^{-8}$&$5.6\times10^{-10}$\\
\end{tabular}
\caption{Nuclear matrix elements $\Phi$ for dimension-$7$, $9$ and $11$ operators.\label{NMEs}}
\end{table}

From this point on, we will assume only one operator is dominant at a time and
ignore any interference between them.  For
the dimension-$7$ operators, the half life, in years, for $\vbb$ is given as:
\bea
\tau^{-1}& =& \frac{v^6}{128 \Lambda^6}\left[{\cal C}_{7,i}\left|\lambda_{7,i}
\right|^2\right]\Phi_{7,i}  \mbox{ yr}^{-1}.
\eea
Here ${\cal C}_{7,i}$ are prefactors that depend on the operator's Lorentz structure, and
$\Phi_{7,i}$ is the nuclear matrix element for the decay.
For dimension-$9$ operators, the rate is given by:
\bea
\tau^{-1}& =& \frac{m_p^2 v^8}{64 \Lambda^{10}}\left[{\cal C}_{9,i}\left|\lambda_{9,i}
\right|^2\right]\Phi_{9,i}  \mbox{ yr}^{-1}.
\eea
Finally, for the dimension-$11$ operators, the rate is:
\bea
\tau^{-1}& = & \frac{m_p^2v^{12}}{256\Lambda^{14}}\left[{\cal C}_{11,i}\left|\lambda_{11,i}
\right|^2\right]\Phi_{11,i}  \mbox{ yr}^{-1}.
\eea

In each of the above, $\lambda$ is the coupling constant of the
operator, $m_p$ is the mass of the proton and ${\cal C}_{i,j}$ are the prefactors given by:
$$
{\cal C}_{7,(1,3,5)}  = 16, \quad {\cal C}_{7,(2,4)}  = 1
$$
\beq
{\cal C}_{9,(1,3,5,7,8,10,12)} = 16,\quad
{\cal C}_{9,(2,4,6)}  = 1, \quad {\cal C}_{9,(9,11)} = 4
\eeq
$$
{\cal C}_{11,(1,2,3,5,7,9,11,13)} = 16,\quad
{\cal C}_{11,(4,6)}  = 4,\quad
{\cal C}_{11,(8,10,12)} =  1
$$

Obviously the strength of an operator's contribution to the rate of
$\vbb$ generally depends on the parameters, $\lambda$ and $\Lambda$.
However there is also a large dependence on the nuclear matrix
elements which describe the complicated physics of the nucleus decaying from (Z,A) to (Z+2,A).
An in-depth discussion on nuclear matrix elements is beyond the scope of this paper but
for a more general survey of this problem see Ref.~\cite{Avignone:2007fu}.
These matrix elements depend crucially on the Lorentz structure of the operator and
the choice of nucleus.
We will use matrix elements for $^{76}$Ge derived in \cite{Pas:1999fc,Pas:2000vn}
and tabulated in \cite{Choi:2002bb}.
The nuclear matrix elements, $\Phi_{\lambda_{i,j}}$ for dimension-$7$, $9$ and $11$ are found in Tables~\ref{NMEs}.

The problem now arises that if we detect a signal of $\vbb$ we do not know whether its source is a
small Majorana neutrino mass or a new LNV operator. If the scale probed by $\vbb$ is higher than a few TeV,
no currently planned accelerator will be able to test the hypothesis of hard LNV operators directly.
But if the scales probed by $\vbb$ are of order the TeV scale or lower, there is reason to hope that
we can unravel the question of hard LNV operators and Majorana masses through a combination $\vbb$
measurements and direct searches for new, high-energy sources of LNV.

\begin{figure}
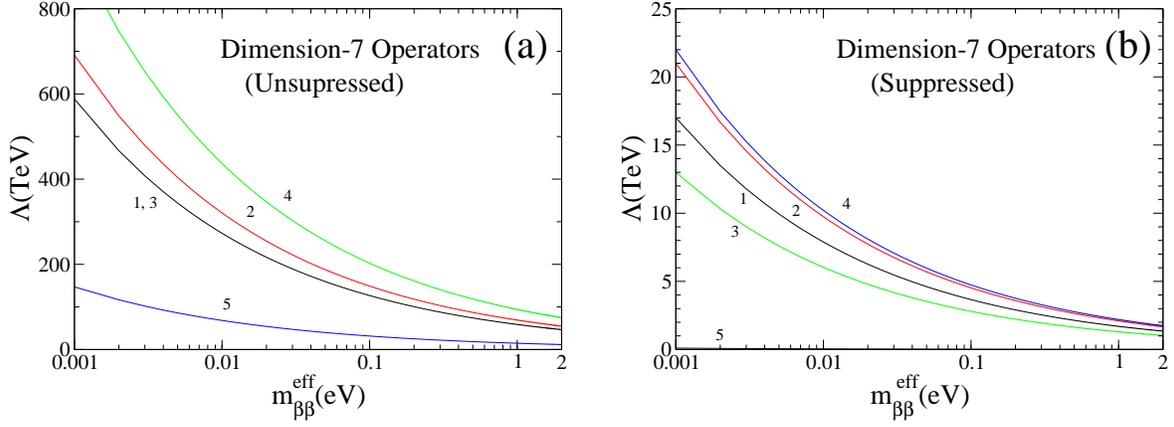

\centering
\includegraphics[scale = .30]{lambdacorgraph.eps}~~~~
\includegraphics[scale = .30]{lambdasupcorgraph.eps}
\caption{Scale of new physics ($y$-axis) being probed by the $\vbb$ experiments when
sensitive to a given effective neutrino mass ($x$-axis), for dimension-$7$ operators.
(a) shows the scale using naive dimensional analysis, while (b) shows the scale after including
constraints from MFV. Note the change in scale on the $y$-axis.\label{ds}}
\end{figure}

\begin{figure}
\centering
\includegraphics[scale = .30]{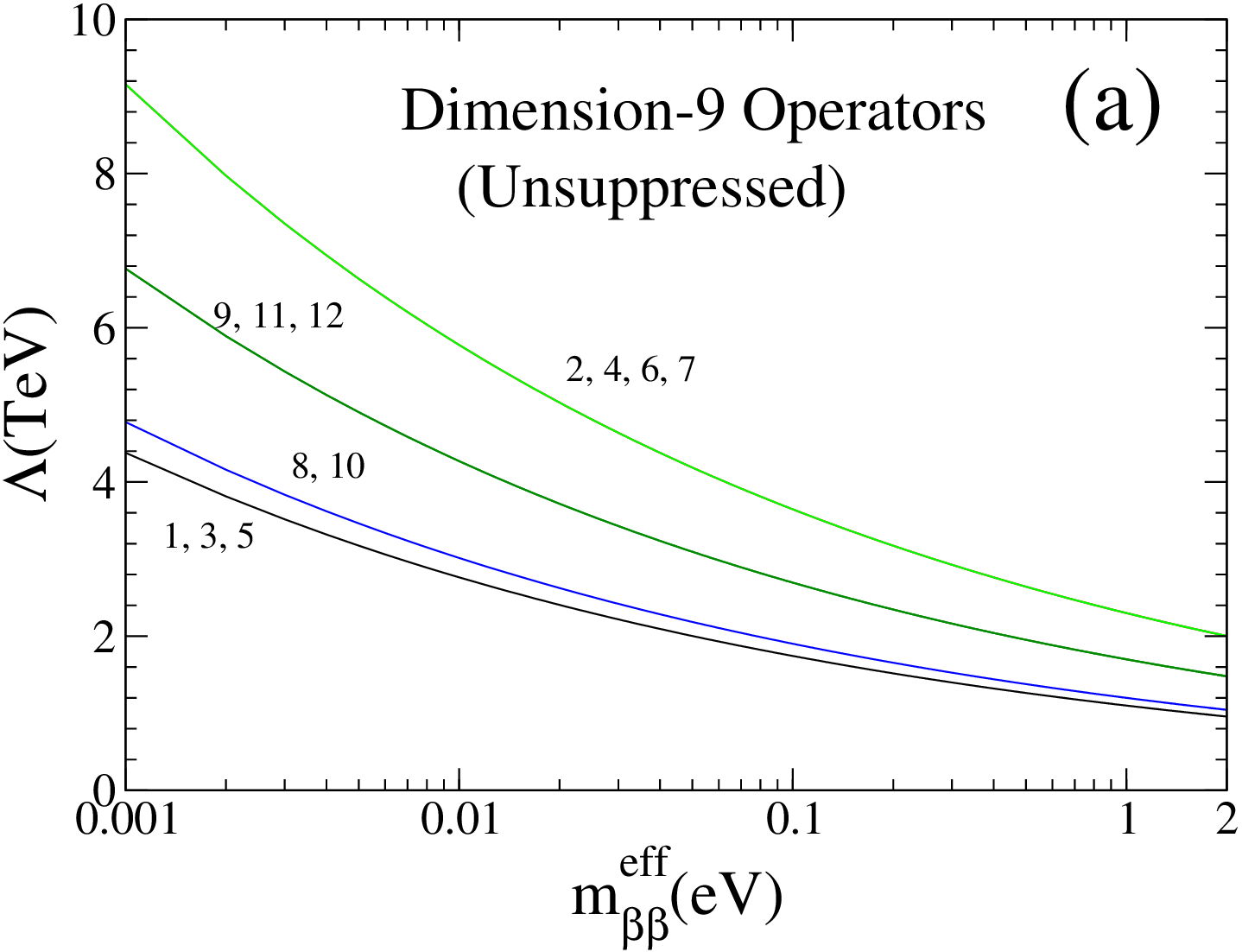}~~~~
\includegraphics[scale = .30]{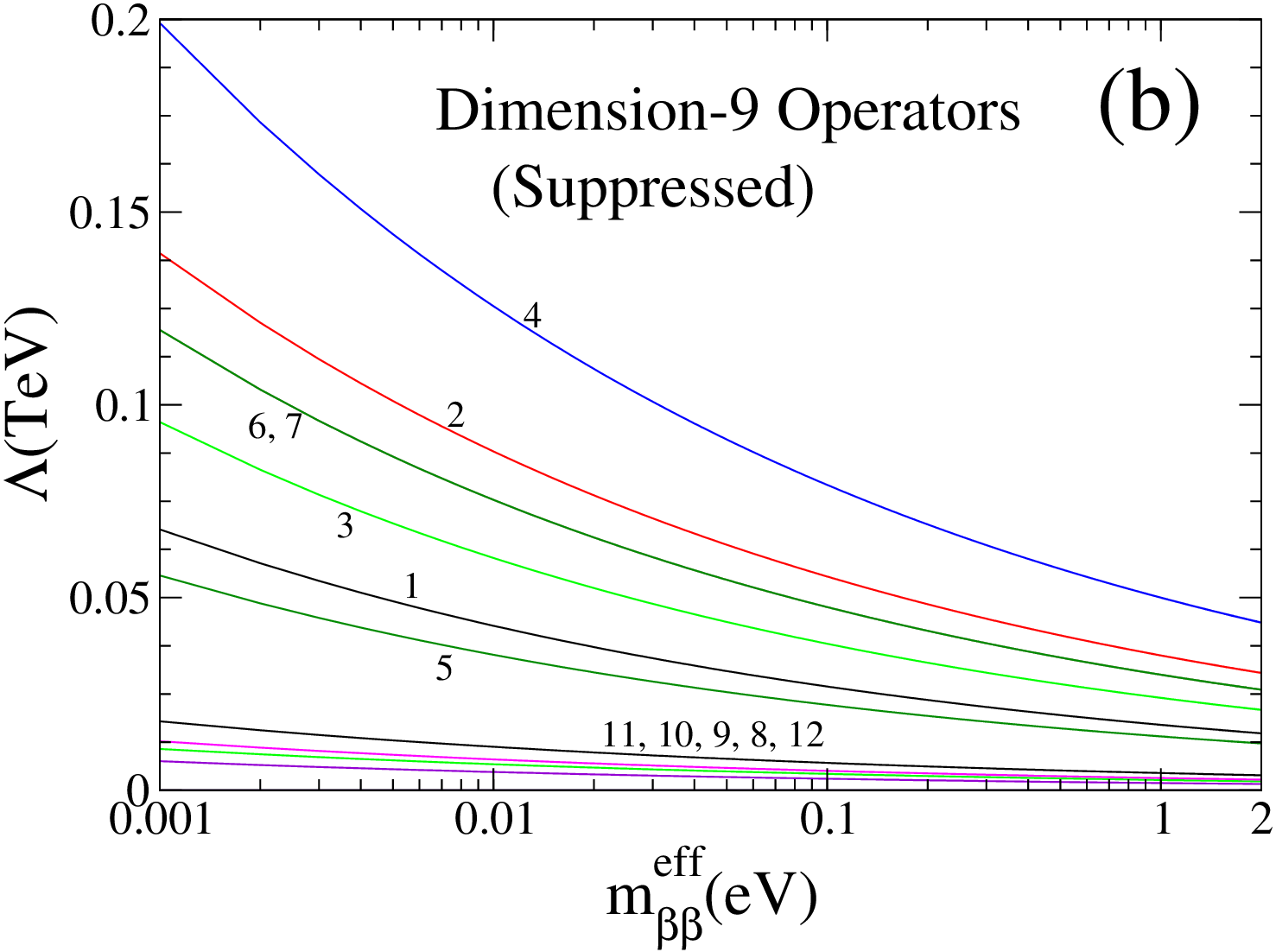}
\caption{Scale of new physics ($y$-axis) being probed by the $\vbb$ experiments when
sensitive to a given effective neutrino mass ($x$-axis), for dimension-$9$ operators.
(a) shows the scale using naive dimensional analysis, while (b) shows the scale after including
constraints from MFV. Note the change in scale on the $y$-axis.\label{dn}}
\end{figure}

\begin{figure}
\centering
\includegraphics[scale = .30]{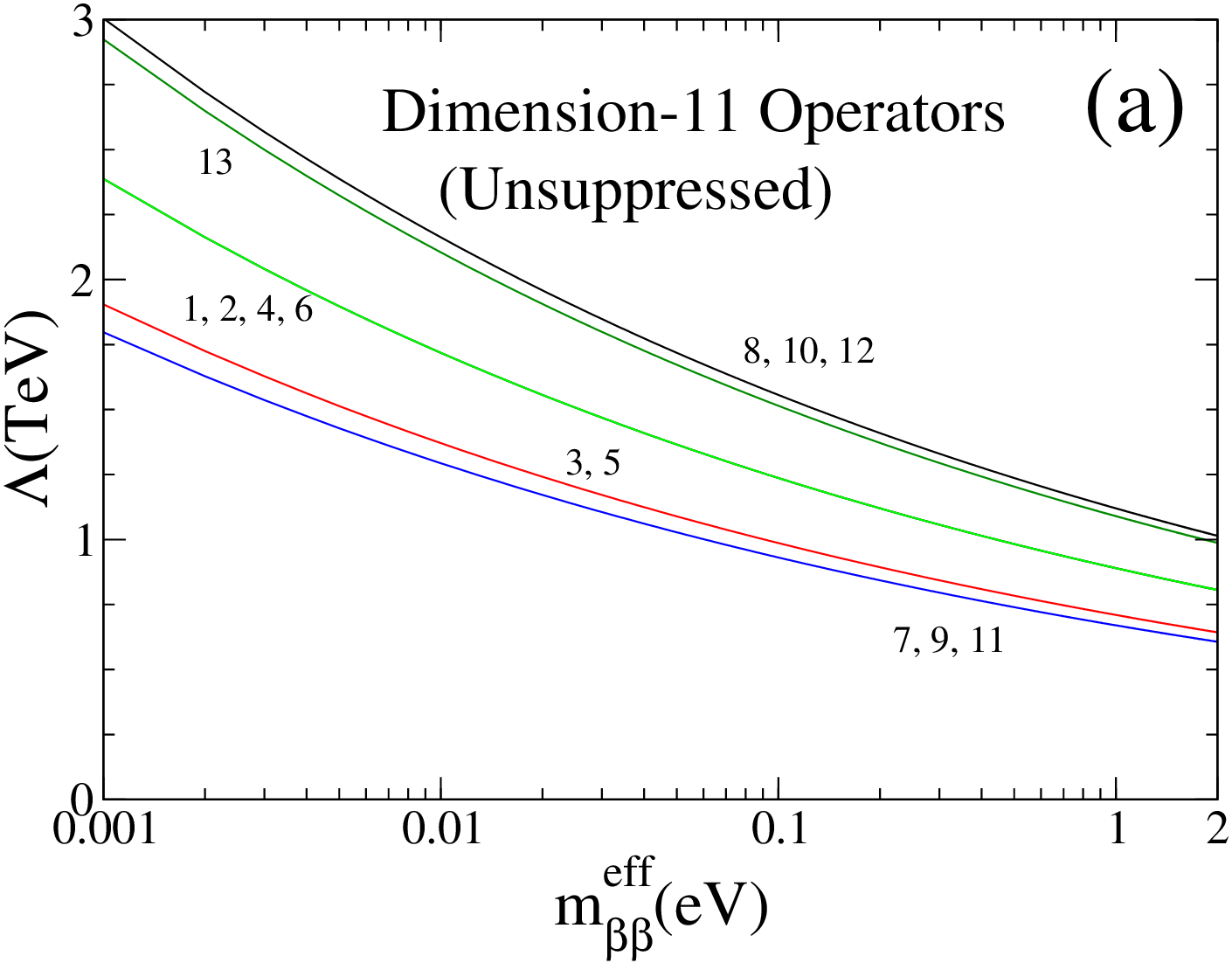}~~~~
\includegraphics[scale = .30]{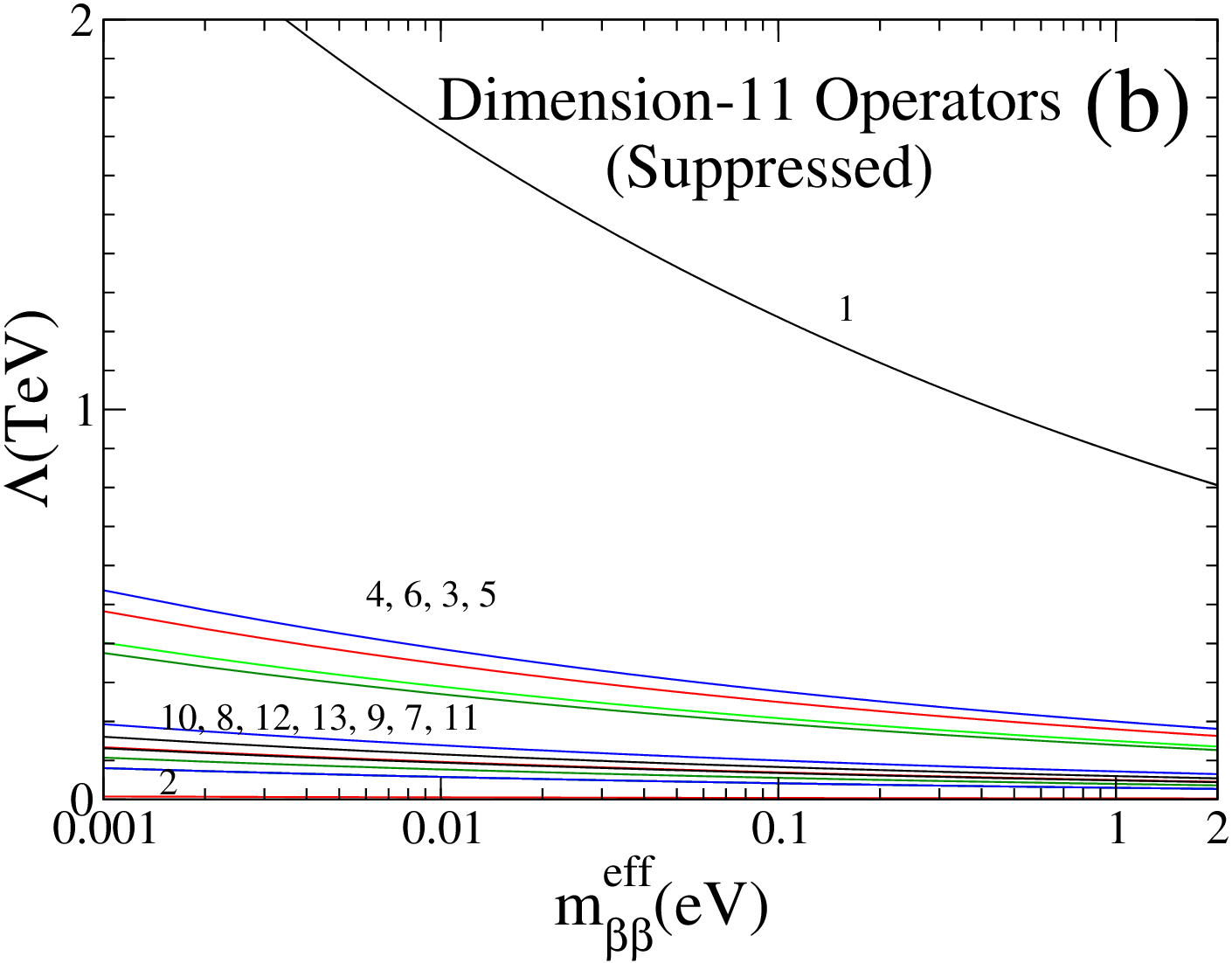}
\caption{Scale of new physics ($y$-axis) being probed by the $\vbb$ experiments when
sensitive to a given effective neutrino mass ($x$-axis), for dimension-$11$ operators.
(a) shows the scale using naive dimensional analysis, while (b) shows the scale after including
constraints from MFV. Note the change in scale on the $y$-axis.\label{de}}
\end{figure}

So the question of which scales are being probed by $\vbb$ experiments is crucial.
In Fig.~\ref{ds}(a), ~\ref{dn}(a) and \ref{de}(a)
we show the naive scales being probed as a function of neutrino mass, for
operators of dimension $7$, $9$ and $11$ respectively.  By ``naive" we mean the scale
obtained by assuming all $\lambda\sim{\cal O}(1)$. In the graphs, the $x$-axis
is the effective neutrino mass, $\mbbeff$ measured in $\vbb$ experiments and $\Lambda$ is the energy scale
at which these operators would need to appear in order to generate the signature
correlating to that effective mass.  
The observable $\mbbeff$ is the value of $m_{\beta\beta}$ extracted by a measurement of $\vbb$. For example, for $^{76}$Ge we simply invert Eq.~(\ref{eq:dim5rate}):
$$
\mbbeff = \left(\frac{2.2\times 10^{30}\,\mbox{yr}}{\tau}\right)^{1/2} (1\times 10^{-3}\ev).
$$
The graphs show that for the
current reach of $\vbb$ experiments($\mbbeff\approx 0.1$ eV),
all the operators could appear at energy scales anywhere from the TeV scale ($d=11$) all the way
to the hundreds of TeV ($d=7$).  These
scales are incredibly high in some cases, too high for the LHC to probe. Thus we should not expect
the LHC to resolve whether a $\vbb$ signal comes from a Majorana mass, or some new physics.

However these energy scales, large as they are, still fall within a very interesting range.  New physics models at scales
$10^{2-3}\tev$ are severely constrained by quark flavor physics.  The Kaon in particular
requires that the scale of new, flavor-violating or CP-violating physics, lie above
several hundred to several thousands of TeV. To get around these constraints,
new physics models often invoke the minimal flavor violation hypothesis.

\section{Minimal Flavor Violation}

Despite the success of the SM at scales below a TeV, there are some sectors in which
the SM is successful at describing physics well above a TeV. If one supposes that there
exists a new source of physics that couples to quarks, then that new physics must generically
occur at scales in the 100's to 1000's of TeV. That is, the SM explanation
for flavor mixing and CP violation through the Yukawa interactions is so successful that
any new interactions which would upset this structure must sit several order above the TeV scale.
It is possible to build models of new physics coupling to the quark sector at lower scales,
but these models must be non-generic. In particular, the only broad class of models which fits this
description are those that exhibit minimal flavor violation.

Minimal flavor violation (MFV) can best be summarized as the condition that flavor and
CP violation in the quark sector comes entirely from the CKM matrix, which is to say, the SM
Yukawa couplings. As it applies to models of new physics, it is the statement that all new couplings
in the flavor sector must either preserve flavor (like a $Z$ with universal couplings) or break it
in ways mimicking the usual Yukawas.
To formulate MFV, we first notice that the SM obeys a global flavor symmetry $U(3)^5$ that is only
broken by the Yukawa couplings.  This symmetry cane be decomposed as
\begin{equation}
SU(3)_Q^3\times SU(3)_L^2\times U(1)_B\times U(1)_L \times U(1)_Y\times U(1)_{PQ}\times U(1)_{E_R},
\end{equation}
where $SU(3)_Q^3$ contains the flavor symmetries for
the left-handed quarks, right-handed up quarks and right-handed
down quarks, and $SU(3)_L^2$ contains the flavor for left-handed and right-handed leptons.
The remaining $U(1)$ factors are associated with the conservation of baryon number, lepton number,
global hypercharge, Peccei-Quin charge and a charge associated with right=handed charged leptons,
none of which will play any role in the rest of our discussion. Because we are assuming that there
is new physics which violated lepton number, we will assume that the $SU(3)_L^2$ (and $U(1)_L$)
are badly broken at or below the TeV scale, but quark flavor constraints do not allow us to make the same
assumption for the $SU(3)_Q^3$.

However, the $SU(3)_Q$ components of this flavor group are broken
by the Yukawas, which can be treated as spurions with the
following charges under $SU(3)_{Q_L}\times SU(3)_{U_R}\times SU(3)_{D_R}$:
\bea
Y_U &=& (3,\bar{3},1),\nonumber\\
Y_D &=& (3,1,\bar{3}).
\eea
These assignments leave the SM mass terms, $\bar{Q} Y_U u_R H+ \bar{Q}Y_Dd_R\tilde{H}$ invariant under the global
$SU(3)^3_Q$ symmetry. When applied to a model of new physics, MFV requires that the {\it only}\/ source
of quark flavor violation be the Yukawas. In an effective theory, this is easy to enforce by
inserting Yukawas as spurions wherever necessary in order to make the new physics
$SU(3)_Q^3$-invariant.

Among the operators that generate $\vbb$, most violate this quark flavor symmetry.
For example, the dimension-$7$ operator,
$\lambda_{7,1} (\nu_e e)(u d^c)$ transforms under $SU(3)_Q^3$ as $(3,1,\bar{3})$. (Recall the $u$
originated from a quark doublet and the $d^c$ from a singlet). MFV then requires an
insertion of $Y^{\dagger}_D$ to remain flavor neutral. Nearly
all the other operators acquire similar suppressions. Table~\ref{suppressiontable}
summarizes the insertions necessary to preserve MFVamong the $\vbb$ operators.

\begin{table}[t]
\centering
\begin{tabular}{c|c|c}
$\lambda_{7,1},\lambda_{7,2}$&$\lambda_{7,3},\lambda_{7,4}$&$\lambda_{7,5}$\\
\hline
$Y^{\dagger}_D$&$Y_U$&$Y_U Y^{\dagger}_D$\\
\hline
$y_d$&$V_{ud}y_u$&$V_{ud}y_uy_d$\\
\end{tabular}

\vspace{0.5cm}

\begin{tabular}{c|c|c|c|c|c}
$\lambda_{9,1},\lambda_{9,2}$&$\lambda_{9,3},\lambda_{9,4}$&$\lambda_{9,5},\lambda_{9,6},\lambda_{9,7}$&
$\lambda_{9,8},\lambda_{9,9}$&$\lambda_{9,10},\lambda_{9,11}$&$\lambda_{9,12}$\\
\hline
${Y^{\dagger}_D}^2$&$Y_U^2$&$Y_U Y^{\dagger}_D$&
${Y^{\dagger 2}_D} Y_U$&$Y_U^2 Y^{\dagger}_D$&${Y^{\dagger 2}_D} Y_U^2$\\
\hline
$y_d^2$&$V_{cd}V_{us}y_cy_u$&$V_{ud}y_uy_d$&
$V_{ud}y_uy_d^2$&$V_{cd}V_{us}y_cy_uy_d$&$V_{cd}V_{us}y_cy_uy_d^2$\\
\end{tabular}

\vspace{0.5cm}

\begin{tabular}{c|c|c|c|c|c|c}
$\lambda_{11,1}$&$\lambda_{11,2}$&$\lambda_{11,3},\lambda_{11,4}$&$\lambda_{11,5},\lambda_{11,6}$&
$\lambda_{11,7},\lambda_{11,8}$&$\lambda_{11,9},\lambda_{11,10}$&$\lambda_{11,11},\lambda_{11,12},\lambda_{11,13}$\\
\hline
$1$&${Y^{\dagger 2}_D} Y_U^2$&${Y^{\dagger}_D}$&$Y_U$&${Y^{\dagger 2}_D}$&$Y_U^2$&
${Y^{\dagger}_D} Y_U$\\
\hline
$1$&$V_{cd}V_{us}y_cy_uy_d^2$&$y_d$&$V_{ud}y_u$&
$y_d^2$&$V_{cd}V_{us}y_cy_u$&$V_{ud}y_uy_d$\\
\end{tabular}
\caption{MFV suppressions of the $\vbb$ operators.\label{suppressiontable}}
\end{table}

In order to determine the precise suppression of any particular $\vbb$ operator, we need to
evaluate the Yukawa insertions. For that, we need to choose the flavor basis. For example,
we could choose to work in a basis in which $Y_D$ is diagonal, in which case
$Y_U = \vckm Y^{diag}_U$, with $Y^{diag}_U = \mbox{diag}(y_u,y_c,y_t)$; conversely
we could choose a basis in which the $Y_U$ is diagonal and the CKM matrix multiplies $Y_D$.
For experimental observables, this basis dependence drops out, but since the Yukawas
are not themselve observable, the suppressions required by MFV are basis dependent.
Luckily, it matters little which basis you choose; the suppressions suffered by each operator will usually
change only by powers of $V_{ud}\sim 1$, which is meaningless in this kind of analysis. But
we will choose the first basis defined above in order to extract specific numbers.  Then
for example, the $\lambda_{9,4}$ coupling picks up a $Y_U^2$ suppression, which can be
rewritten as $\vckm Y^{diag}_U\vckm Y^{diag}_U$; the (1,1) component of this matrix is
dominated numerically by the term $V_{cd} V_{us}y_uy_c$. The piece which dominates the suppression
for each operator is also shown in Table~\ref{suppressiontable}.

It is worth noting that while most of the operators break the chiral symmetry
embedded in the full $SU(3)^3_Q$ flavor symmetry, a few of the operators (corresponding to
$\lambda_{7,5}$, $\lambda_{9,7}$ and
$\lambda_{9,12}$) don't break the chiral symmetry. This is because the
operators correspond to pure right-handed currents. Models with right-handed
currents (such as left-right models) would almost certainly break the flavor
symmetry and would therefore not be MFV; such models are beyond the scope of this paper. But as long as MFV does hold,
we can assign suppressions to these operators using the full $SU(3)^3_Q$.

The effect of the Yukawa suppressions is that couplings which are usually assumed to be ${\cal O}(1)$
are in fact highly suppressed,
by as much as $10^{-18}$ in the case of $\lambda_{9,12}$. Then in order to get an observable rate
in $\vbb$, the relevant energy scale $\Lambda$ must be much closer to the weak scale, than the naive dimensional analysis from the previous section.

In each of the three Figures~\ref{ds}, \ref{dn} and \ref{de}, we have taken the naive energy scaling
in the (a) figure and replaced it in (b) with a scaling that accounts for the MFV Yukawa suppressions.
As expected, we see a significant drop in the energy scales that can be probed in $\vbb$ experiments.
As an example, consider an experiment capable of probing $\vbb$ down to an effective mass $\mbbeff$
of $0.1\ev$. For the dimension-7 operators (see Fig.~\ref{ds}) without MFV constraints, such an experiment
is probing physics at the 100 to $200\tev$ energy scale. With the MFV constraints, the relevant scale is
reduced to 3 to $5\tev$. (The right-handed vector operator corresponding to $\lambda_{7,5}$ becomes so impotent
that it comes in only at scales around $10\gev$ -- presumably we would have seen evidence of this new physics
long before now.)
Sitting at $\Lambda\sim\,$few TeV means that the LHC will not be able to rule out the possibility of
new hard LNV interactions, but it brings the physics into a range at which the LHC has a shot.

For the dimension-$9$ operators, compared in Fig.~\ref{dn}, inclusion of the MFV Yukawa suppressions
means that current generation $\vbb$ experiments are not probing $\Lambda$ in the range of 2 to $4\tev$,
but rather at scales below $150\gev$! This means that LEP already had a fair chance at finding this LNV,
and certainly an LHC or linear collider should be able to do it.

Finally, we have the dimension-$11$ operators in Fig.~\ref{de}.  In order to be observed in $\vbb$, the
dim-$11$ operators always require a relatively light cutoff, usually 1 to $2\tev$.
Once they are suppressed by the MFV Yukawa corrections, these cutoffs mostly fall to a few hundred GeV,
making them easy to access directly, if they aren't already ruled out.
Among these dim-11 operators there is one, corresponding to
$\lambda_{11,1}$, that is not suppressed at all because it is built from purely left-handed
vector currents. This one operator is then the only dim-11 operator sensitive to physics in the
TeV range in $\vbb$ experiments.

Overall, the inclusion of the MFV-required Yukawa suppressions does force the energy scales to
which $\vbb$ is sensitive to fall sharply. Many operators which were possible sources of $\vbb$
at planned and ongoing experiments now appear to be wholly unimportant as they have been suppressed
beyond relevance. Other operators have been suppressed into a range of a few hundred GeV where the
LHC (or a linear collider) might reasonably probe the physics of LNV directly. For a small subclass
of operators, all dim-7 except one at dim-11, there still remains the possibility that they could
generate observable $\vbb$ while being undetectable to the LHC or the next linear collider. For such
operators, a higher energy lepton machine, such as a CLIC, may be the only way to probe for direct sources of LNV.

\section{Induced Neutrino Masses}

Until now we have ignored a potentially important effect: neutrino masses
induced by loops involving the LNV
operators that contribute to $\vbb$. These operators can generate sizeable Majorana
neutrino masses, and if these induced masses are large enough, their effect on the $\vbb$
rate will overshadow the direct effects of the operators themselves, making it impossible
to pull out the effects of the higher-dimension operators. However, the tendency of these
operators to generate neutrino masses is not confined to the $\nu_e$ (where the $\vbb$ sensitivity lies).
These operators also generate masses for $\nu_\tau$ and $\nu_\mu$, all of which are constrained by
neutrino mixing measurements. A very nice analysis of this effect was done by DeGouvea and
Jenkins~\cite{de Gouvea:2007xp}, where it was assumed that higher-dimension operators came in with
coefficients of $O(1)$. In this section we repeat their analysis adding the constraints
implied by  MFV. But we only consider here the operators that lead to direct $\vbb$.

In order to generate neutrino masses, we must close off the external fermion lines in
the $\vbb$ operators, which requires us to work at multiple loops. We will assume that
no new physics states occur inside the loops, evaluating the loop integrals only up to
the cutoff scale $\Lambda$. Most of the loops are divergent, bringing in powers of $\Lambda$
that cancel the $1/\Lambda^{d-4}$ suppression inherent in a dim-$d>4$ operator.

If one only considers the operators exactly as they contribute to $\vbb$ (that is, with $u$- and $d$-quarks on external lines), one finds that the 
induced neutrino masses are smaller than the effective masses, which means that any observation of $\vbb$
may be due to hard LNV operators rather than Majorana mass terms~\cite{Choi:2002bb}.
This is because many of the quark loops can only be closed using $u$- or $d$-quark mass
insertions, generating suppressions in the induced mass with powers of $(m_{u,d}/\Lambda)$.

In a generic model, this is the best one can do. For example, consider the $\vbb$-generating
operator $(\nu_e e)(ud^c)$ and its flavor cousin $(\nu_e e)(tb^c)$. Both operators induce a
Majorana mass for the neutrino, and if both came in at $O(1)$, the second operator would
induce a mass which is $m_b/m_d$ times larger than that of the first operator. But in a
generic model there is no need to assume that both operators have the same strength. Thus we
could simply ignore the possibility of the second operator when considering the effects of the first. 

In MFV models, however, there is a strong correlation between these operators. Both come
from the electroweak operator $(LL)(Qd^c)H$ and require a single insertion of $Y_D$ in
order to preserve the quark flavor symmetries. For the first generation quark current,
this is a large suppression of $O(y_d)$, while for the third generation quark current the
suppression is much smaller, $O(y_b)$. Since both operators generate a mass for the same
neutrino, it is clear that the latter operator will dominate. 

There is a similar argument for the operators $(\nu_e e)(ud^c)$ and $(\nu_\tau \tau)(ud^c)$. One might expect the existence of the first to imply the second, but in the present analysis we make no assumptions about the lepton flavor symmetry, and so have no method for correlating the coefficients of these two operators. Thus in our calculation of induced neutrino masses, we only consider the effects of the $(\nu_e e)$-type operators, since these correspond directly to what is observable in $\vbb$. There are arguments that the coefficients of $(\nu_e e)$ operators should be of the same order as those for the $(\nu_\tau \tau)$ operators (\eg, the large $\nu$-mixing implied by the atmospheric data); we discuss that possibility briefly below.

We have gone through each operator which generates $\vbb$ at dimension $7$ through $11$
and evaluated (very roughly) the size of the neutrino mass induced by each operator. Our
results follow those of Ref.~\cite{de Gouvea:2007xp}, except for the $\lambda_{9,12}$ operator,
which does not appear in that paper. That operator is composed entirely of right-handed fields, and generates a neutrino mass of order $y_t^2 y_b^2 y_e^2/(16\pi^2)^4\times (v^2/\Lambda)$.

\begin{figure}[t]
\centering
\includegraphics[scale = .45]{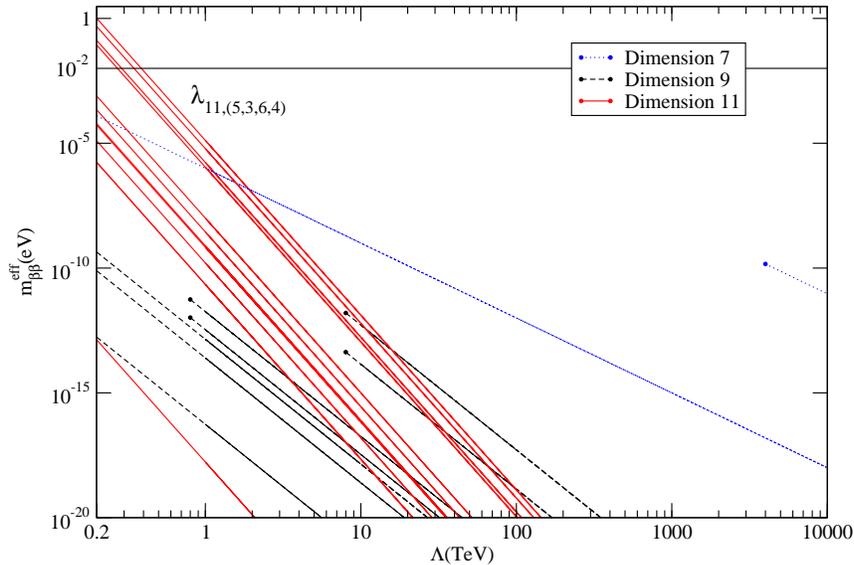}
\caption{Lower limits on the energy scale for which the operator induces a neutrino mass
less than $0.05\ev$, versus the effective mass ($\mbbeff$) each operator would fake
in $\vbb$ decays. The two labeled operators, $\lambda_{11,3}$ and $\lambda_{11,4}$, are able
to produce observable rates in $\vbb$ directly while keeping $m_\nu\lsim 0.05\ev$. Each line
ends at the scale $\Lambda_0$ below which it would produce $m_\nu\gsim 0.05\ev$. The region above the solid, horizontal line at $\mbbeff=10^{-2}$ is the range in which current and planned $\vbb$ experiments will be sensitive.
\label{vbbbig}}
\end{figure}

In Fig.~\ref{vbbbig}, we show the results of this analysis. Each operator has a lower
cutoff $\Lambda_0$ below which it will generate a neutrino mass above $0.05\ev$. We take
$0.05\ev$ to be the upper bound on all neutrino masses, assuming there to be no fine-tuned
cancellations in the interpretation of the atmospheric neutrino data. If $\Lambda_0$ is
below $200\gev$, we assume that the induced masses are all too small to be of interest;
otherwise we only evaluate that operator's contributions to $\vbb$ at scales above $\Lambda_0$.

In this figure we see that only four operators satisfy both the upper bound on $m_\nu$,
and generate a rate for $\vbb$ which will be observable in the coming generation of experiments.
Those four operators are all dimension-$11$ ($\lambda_{11,(3,4,5,6)}$) and have the form:
$$
(e\sigma^{\mu}\bar{e}^c)(q\sigma^{\nu}\bar{q})(q \Gamma_{\mu\nu} q^c)
$$
with $\Gamma_{\mu\nu} = \delta_{\mu\nu}$ (scalar) or $\sigma_{\mu\nu}$ (tensor). The two quark currents imply only one $Y_{U,D}$ suppression, since the first is actually the usual weak current. The $(e\sigma^\mu \bar e^c)$ requires a single factor of $y_e$ in order to generate a neutrino mass. All other dim-11 operators either require additional quark flavor suppressions (which suppresses their rates for $\vbb$) or induce neutrino masses with fewer Yukawa or loop suppressions (which forces their cutoffs to be large). Only the particular combination represented by the four operators $\lambda_{11,(3,4,5,6)}$ leads to a viable $\vbb$ signal consistent with neutrino mass bounds.

If, in addition, we impose some form of lepton flavor universality, so that the existence of $(e\sigma^\mu \bar e^c)$ operators implies $(\tau\sigma^\mu \bar\tau^c)$ operators of the same size, then two of these four operators ($\lambda_{11,(5,6)}$) are no longer viable. The cutoff for these two operators is pushed high enough ($\Lambda_0\approx 2\tev$) to suppress the rate for $\vbb$ below the range of interest.

Our conclusion is that a signal in the current or next generation of $\vbb$ experiments could be either a measurement of a Majorana mass, or the observation of one of these four dim-11 operators, under the assumption of MFV. If the signal does come from one of the dim-11 operators, then the cutoff implied is below a TeV, and so new physics should be in the range accessible to the LHC or a linear collider.

This type of analysis
contains several assumptions apart from MFV, which we have noted in passing.  For one thing, we assume
that there are no large, nearly-degenerate neutrino masses. If it were the case that
$\Delta m^2_\nu\ll m^2_\nu$, then the scale $\Lambda_0$ would come down for every operator,
bringing more into the experimentally viable range. This could have a particularly profound effect on some of the dim-7 operators, which are very good at generating $\vbb$ were it not for the large neutrino masses that they induce.

Second, we have assumed that all operators are generated by tree-level effects in the ultraviolet theory, so that there are no loop suppressions in their coefficients. It is easy to imagine that these operators are loop-induced, and so
their coupling constants have additional $1/16\pi^2$ suppressions. These loop factors in the couplings would severely suppress the induced neutrino masses, lowering the cutoff scale $\Lambda_0$. But these same loop factors would have a much smaller effect on the scale $\Lambda$ probed by $\vbb$, since $\Lambda$ itself only scales with $\mbbeff$ (and thus the couplings) as the $1/(d-4)$ power.
This may allow one to construct models in which the neutrino mass bound is satisfied while still generating LNV operators
which could be probed by $\vbb$ experiments.

\section{Conclusion}

The question of whether neutrino masses are Dirac or Majorana is one of the most important,
and most difficult, questions remaining in the neutrino sector.
Many experiments have looked for, and will continue to look for neutrinoless double-$\beta$ decay in order to distinquish between the two cases.
However, such a search raises its own questions, as
new LNV operators may also generate $\vbb$, and it will be difficult to determine from
which source such an experimental signal arises. And because $\vbb$ appears
to probe scales far above those accessible at colliders, 
a definitive answer may be far off, requiring measurements of $\vbb$ in multiple nuclei.

However, there is another sector of the SM which is sensitive to such high scales:
the quark CP/flavor sector. In order to avoid new, large flavor-changing neutral currents and CP
violation, nature must preserve the approximate
flavor symmetries of the SM. We took as our assumption the slightly more rigorous statement
that nature is minimally flavor violating in the quark sector, and preserves those flavor
symmetries everywhere except in the usual Yukawa interactions. This assumption then placed an
additional set of constraints and suppressions on new physics operators, even those which are quark
flavor conserving, such as those which generate $\vbb$. We found the effect of these MFV suppressions
is to bring the scale probed by the current and next generation of $\vbb$ experiments into
the same approximate range (less than a few TeV) that will be probed by the LHC.

We then examined the possibility that these same operators induce neutrino masses
inconsistent with our current understanding of the neutrino
mass hierarchy. In particular, we required that the $\vbb$-generating operators and their
flavor cousins not generate any neutrino mass above $0.05\ev$. This constraint effectively
killed most of the dim-7, 9 and 11 operators, leaving only four dim-11 operators as viable
sources of $\vbb$ in upcoming experiments. And even for these four operators, new physics
must be sitting at scales below a TeV.

Therefore, if $\vbb$ were to be observed in the near future, and no new physics is
observed at the LHC, minimal flavor violation would seem to require that the observation is
indeed the measurement of a Majorana neutrino mass. Considering that these constraints
came from quark flavor changing, it is remarkable that they have had such a profound effect on processes of lepton number non-conservation.

\section*{Acknowledgements}
We would like to think K.S.~Babu, J.~Engel and I.~Gogoladze for helpful conversations.
This work was partially supported by the National Science Foundation under grant PHY-0355066 and by the Notre Dame Center for Applied Mathematics.


\begin{thebibliography}{99}
\bibitem{Fukuda:2001nk}
  S.~Fukuda {\it et al.}  [Super-Kamiokande Collaboration],
  Phys.\ Rev.\ Lett.\  {\bf 86}, 5656 (2001)
  [arXiv:hep-ex/0103033].

\bibitem{Apollonio:1999ae}
  M.~Apollonio {\it et al.}  [CHOOZ Collaboration],
  Phys.\ Lett.\  B {\bf 466}, 415 (1999)
  [arXiv:hep-ex/9907037].

\bibitem{Ahn:2002up}
  M.~H.~Ahn {\it et al.}  [K2K Collaboration],
  Phys.\ Rev.\ Lett.\  {\bf 90}, 041801 (2003)
  [arXiv:hep-ex/0212007].

\bibitem{Fukuda:1998mi}
  Y.~Fukuda {\it et al.}  [Super-Kamiokande Collaboration],
  Phys.\ Rev.\ Lett.\  {\bf 81}, 1562 (1998)
  [arXiv:hep-ex/9807003].

\bibitem{Ahmad:2002jz}
  Q.~R.~Ahmad {\it et al.}  [SNO Collaboration],
  Phys.\ Rev.\ Lett.\  {\bf 89}, 011301 (2002)
  [arXiv:nucl-ex/0204008].

\bibitem{KlapdorKleingrothaus:2000sn}
  H.~V.~Klapdor-Kleingrothaus {\it et al.},
  Eur.\ Phys.\ J.\  A {\bf 12}, 147 (2001)
  [arXiv:hep-ph/0103062].

\bibitem{Baudis:1999xd}
  L.~Baudis {\it et al.},
  Phys.\ Rev.\ Lett.\  {\bf 83}, 41 (1999)
  [arXiv:hep-ex/9902014].

\bibitem{Aalseth:2002rf}
  C.~E.~Aalseth {\it et al.}  [IGEX Collaboration],
  Phys.\ Rev.\  D {\bf 65}, 092007 (2002)
  [arXiv:hep-ex/0202026].

\bibitem{Elliott:2002xe}
  S.~R.~Elliott and P.~Vogel,
  Ann.\ Rev.\ Nucl.\ Part.\ Sci.\  {\bf 52}, 115 (2002)
  [arXiv:hep-ph/0202264].

\bibitem{KlapdorKleingrothaus:2001ke}
  H.~V.~Klapdor-Kleingrothaus, A.~Dietz, H.~L.~Harney and I.~V.~Krivosheina,
  Mod.\ Phys.\ Lett.\  A {\bf 16}, 2409 (2001)
  [arXiv:hep-ph/0201231].

\bibitem{Aalseth:2002dt}
  C.~E.~Aalseth {\it et al.},
  Mod.\ Phys.\ Lett.\  A {\bf 17}, 1475 (2002)
  [arXiv:hep-ex/0202018].

\bibitem{Choi:2002bb}
  K.~w.~Choi, K.~S.~Jeong and W.~Y.~Song,
  Phys.\ Rev.\  D {\bf 66}, 093007 (2002)
  [arXiv:hep-ph/0207180].


\bibitem{D'Ambrosio:2002ex}
  G.~D'Ambrosio, G.~F.~Giudice, G.~Isidori and A.~Strumia,
  Nucl.\ Phys.\  B {\bf 645}, 155 (2002)
  [arXiv:hep-ph/0207036].

\bibitem{Babu:2001ex}
  K.~S.~Babu and C.~N.~Leung,
  Nucl.\ Phys.\  B {\bf 619}, 667 (2001)
  [arXiv:hep-ph/0106054].

\bibitem{de Gouvea:2007xp}
  A.~de Gouvea and J.~Jenkins,
  Phys.\ Rev.\  D {\bf 77}, 013008 (2008)
  [arXiv:0708.1344 [hep-ph]].

\bibitem{Avignone:2007fu}
  F.~T.~Avignone, S.~R.~Elliott and J.~Engel,
  arXiv:0708.1033 [nucl-ex].

\bibitem{Pas:1999fc}
  H.~Pas, M.~Hirsch, H.~V.~Klapdor-Kleingrothaus and S.~G.~Kovalenko,
  Phys.\ Lett.\  B {\bf 453}, 194 (1999).

\bibitem{Pas:2000vn}
  H.~Pas, M.~Hirsch, H.~V.~Klapdor-Kleingrothaus and S.~G.~Kovalenko,
  Phys.\ Lett.\  B {\bf 498}, 35 (2001)
  [arXiv:hep-ph/0008182].





\end{thebibliography}
\end{document}